\documentclass{pasj00}
\draft

\begin{document}
\SetRunningHead{Y. Takeda et al.}{Abundance of Potassium in Metal-Poor Stars}
\Received{2001/10/08}
\Accepted{2002/02/20}

\title{
On the Abundance of Potassium in Metal-Poor Stars
}

\author{Yoichi \textsc{Takeda}}
\affil{Komazawa University, Komazawa, Setagaya, Tokyo 154-8525}
\email{takedayi@cc.nao.ac.jp}
\author{Gang \textsc{Zhao}, Yu-Qin \textsc{Chen}, and Hong-Mei \textsc{Qiu}}
\affil{National Astronomical Observatories, Chinese Academy
 of Sciences, Beijing 100012, P. R. China \\ {\rm and}}
\author{Masahide \textsc{Takada-Hidai}}
 \affil{Liberal Arts Education Center, Tokai University,
 Hiratsuka, Kanagawa 259-1292}

%


\maketitle

\begin{abstract}
Based on extensive statistical-equilibrium calculations,
we performed a non-LTE analysis of the K~{\sc i} 7699 equivalent-width 
data of metal-deficient stars for the purpose of clarifying the behavior 
of the photospheric potassium abundance in disk/halo stars.
While the resulting non-LTE abundance corrections turned out to be
considerably large, amounting to 0.2--0.7 dex, their effect on the
[K/Fe] vs. [Fe/H] relation is not very important, since these 
corrections do not show any significant metallicity dependence.
Hence, we again confirmed the results of previous LTE studies, that
[K/Fe] shows a gradual systematic increase toward a lowered metallicity
up to [K/Fe] $\sim$ 0.3--0.5 at [Fe/H] $\sim$ $-1$ to $-2$,
such as in the case of $\alpha$~elements. 
\end{abstract}

%


\section{Introduction}

Among the elements for which abundances can be spectroscopically
determined in metal-poor stars to investigate the chemical history 
of our Galaxy, our current knowledge about potassium (K; $Z=19$) 
is still very insufficient in the sense that considerable ambiguities
are involved in both the theoretical galactic chemical evolution 
calculation regarding [K/Fe] and the observational aspect
of K abundance determination.

Theoretical investigations on the chemical evolution of this
alkali element still suffer rather large uncertainties.
Based on Woosely and Weaver's (1995; hereinafter WW95) yields
and Salpeter's (1955) Initial Mass Function (IMF),
Timmes et al. (1995) suggested a decreasing trend of [K/Fe]
with a lowering of [Fe/H] (i.e., negative [K/Fe] in the metal-poor 
regime). Even when the Fe yield of WW95 is reduced by a factor of 2
(which they suggested to be more reasonable), their [K/Fe] barely 
exceeds $\sim 0$ (cf. their figure 24). Unfortunately, such a tendency 
apparently contradicted the observational implication 
of supersolar [K/Fe] (just like $\alpha$ elements) in metal-poor stars
(see the next paragraph). This situation was improved by a
recent calculation of Goswami and Prantzos (2000), who also used
the metallicity-dependent yields of WW95 (with a reduction of Fe yield
by a factor of 2), but the more realistic IMF of Kroupa et al. (1993)
(along with their own halo model). Their resulting [K/Fe] shows a
mildly supersolar behavior of [K/Fe]$\sim +0.2$ (cf. figure 7 therein).
However, even this still appears to be somewhat insufficient to account for
the observed [K/Fe], amounting up to $\sim +0.5$. If a more satisfactory
consistency is to be further pursued, the K yield itself would have to 
be adequately adjusted, such as was done by Samland (1998), who used a 
K yield about twice as large as that of WW95 (along with a Fe yield of about 
half that of WW95) and Salpeter IMF, by which he could
reproduce the tendency that [K/Fe] gradually increases from $\sim 0$ 
(at [Fe/H] $\sim 0$) to $\sim +0.3$ (at [Fe/H] $\sim -1$) and
maintains a nearly constant value of [K/Fe] $\sim +0.3$ at
$-3 \ltsim$~[Fe/H]~$\ltsim -1$, being nearly similar to the
observed trend (see below). We should bear in mind, however, that 
such empirical adjustments of the yields are involved in his calculation.

Let us then turn our attention to the observational side.
As far as we know, there have been only two studies which attempted to
investigate the potassium abundance of metal-deficient stars.
Gratton and Sneden (1987a,b) were the first to determine the K abundances 
of disk/halo stars (dwarfs as well as giants) in the metallicity range of 
$-2.5 \ltsim$~[Fe/H]~$\ltsim +0.3$ by using the K~{\sc i} resonance
doublet lines at 7664.87~$\rm\AA$ and 7698.98~$\rm\AA$.
Meanwhile, Chen et al. (2000) recently carried out an extensive 
investigation on the chemical composition of mildly metal-poor 
($-1 \ltsim$~[Fe/H]~$\ltsim 0$) F--G disk dwarfs, and determined 
the abundance of K, being included as one of their target elements, 
by using the K~{\sc i} 7698.98~$\rm\AA$ line.
Interestingly, both of these two studies suggested a gradually increasing
trend of [K/Fe] with a lowering of the metallicity at 
$-1 \ltsim$~[Fe/H]~$\ltsim 0$ (though with a rather large diversity), 
as has usually been referred to by theoreticians in comparing their 
computed predictions with observations.
However, these authors do not appear to be sufficiently confident about 
the results which they obtained, since various difficulties are involved in 
the spectroscopic determination of the K abundance.

It should be kept in mind that only the resonance doublet lines 
(K~{\sc i} 7665 and 7699) are practically available, since other 
subordinate lines (cf. Lambert, Warner 1968) are either too weak to be 
measurable in metal-poor stars, or are located in an unfavorable 
wavelength region ($\lambda \sim 1.2 \; \mu$m).
Unfortunately, however, these resonance lines are not necessarily 
suitable for an abundance determination for the following reasons:\\
--- They occasionally suffer appreciable contaminations
of the telluric O$_{2}$ lines; this blending effect is often more serious
for the K~{\sc i} 7665 line. \\
--- Since they are sufficiently strong to be saturated, the abundance
is appreciably affected by the choice of the microturbulent velocity,
which may be uncertain owing to its possible depth-dependence.\\
--- Similarly, due to their considerable strengths, the abundances are 
more or less influenced by the adopted damping constant, which is
not well established.\\
--- They are considered to suffer a significantly large non-LTE effect; 
for example, according to the calculation of Takeda et al. (1996),
the non-LTE abundance correction for the Sun (G2~V) and Procyon (F5~III--IV)
amounts to $-0.4$~dex and $-0.7$~dex, respectively.\\
--- In addition, because of its low ionization potential, only a very small
number of potassium atoms remain neutral (i.e.,  almost all are in the 
once-ionization stage), which means that the abundance is sensitive to 
the atmospheric model (e.g., a choice of $T_{\rm eff}$).\\
Presumably because of these difficulties, stellar spectroscopists may 
have somewhat hesitated to struggle with the potassium abundance
of metal-poor stars. 

Considering this situation, we decided to visit this problem 
by performing a non-LTE analysis on the observational K~{\sc i} 7699 data 
of the two studies mentioned above, while aiming to elucidate the behavior 
of [K/Fe] in metal-deficient stars, in order to provide theoreticians
with observational information on the galactic [K/Fe] vs. [Fe/H] relation,
while extensive non-LTE calculations were carried out on model 
atmospheres over a wide range of parameters for this purpose. 
This was the primary motivation of the present study.

We describe the adopted observational data in section 2.
A description of our non-LTE calculations is presented
in section 3. Our abundance determination procedure 
is explained in section 4, followed by section 5, where
the possible uncertainties involved in the resulting
abundances are estimated. The results are finally summarized in section 6.

\section{Observational Data}

The main observational data which we adopted were, similarly to those 
used by Chen et al. (2000), the spectra of 21 mildly metal-poor 
F--G stars observed by the Coud\`e Echelle Spectrograph attached to 
the 2.16~m telescope at Beijing Astronomical Observatory.
See Chen et al. (2000) for more details. Figure 1 shows the 
7692--7722~$\rm\AA$ portion of these spectra, the S/N ratio of which
was estimated to be 150--300. Based on these spectra, the equivalent width
of the K~{\sc i} 7699 line was remeasured by one of us (Y.T.), either by 
Gaussian fitting or direct integration, depending on the situation. 
A comparison of such measured
equivalent widths (which are given in table 1) with those used by 
Chen et al. (2000) (though the equivalent-width data for individual stars 
are not published therein) is shown in figure 2, from which we can see 
that the agreement is quite satisfactory. Namely, the observational data of 
these 22 stars (BAO samples) adopted in this study are essentially the
same as those used by Chen et al. (2000).

Then, we also invoked the K~{\sc i} 7699 equivalent-width data of
24 metal-poor dwarfs and giants published by Gratton and Sneden (1987b),
which were used in Gratton and Sneden's (1987a) analysis. 
(Their K~{\sc i} 7665 data were not used, since we tried to make a
consistent comparison.) These data are also presented in table 1. Note that two stars, 
HD~34411 and HD~142373, are common to both samples.

Regarding the K~{\sc i} 7699 equivalent width for the Sun, which was 
used as the reference standard, we measured it on the solar flux spectrum
published by Kurucz et al. (1984; cf. figure 1) by the direct-integration 
method, and obtained 169.9~m$\rm\AA$, which we eventually adopted. 
This value was further checked on the Moon spectrum observed at BAO. 
Although the K~{\sc i}~7699 line profile on this BAO spectrum shows a 
slight asymmetry in the damping wing (cf. figure 1), and is thus 
comparatively less suitable for an accurate measurement, we confirmed that 
the resulting equivalent-width (Moon) is in fairly good agreement 
(to within a few m$\rm\AA$) with our adopted value.

\section{Statistical Equilibrium Calculations}

The procedures of our non-LTE calculations for neutral potassium
were the same as that described in Takeda et al. (1996), which
should be consulted for details. We only mention here that
the H~{\sc i} collision rates in rate equations were drastically 
suppressed to a negligible level by multiplying the classical rates 
by a factor of $10^{-3}$  ($h=-3$) according to the consequence of 
Takeda et al. (1996).

Since we planned to make our calculations applicable to stars 
from near-solar metallicity (population I) down to very low 
metallicity (extreme population II) at late-F through early-K 
spectral types in various evolutionary stages 
(i.e., dwarfs, subgiants, giants, and supergiants),
we carried out non-LTE calculations on an extensive grid of 
one hundred ($5 \times 5 \times 4$) model atmospheres resulting from 
combinations of five $T_{\rm eff}$ values 
(4500, 5000, 5500, 6000, 6500 K), five $\log g$ values 
(1.0, 2.0, 3.0, 4.0, 5.0), and four metallicities (represented by [Fe/H])
(0.0, $-1.0$, $-2.0$, $-3.0$).
As for the stellar model atmospheres, we adopted Kurucz's (1993) ATLAS9 
models corresponding to a microturbulent velocity ($\xi$) of 2~km~s$^{-1}$.

Regarding the potassium abundance used as an input value in non-LTE 
calculations, we assumed 
$\log\epsilon_{\rm K}^{\rm input}$ = 5.12 + [Fe/H] +[K/Fe],
where [K/Fe] = 0.0 for the solar metallicity models ([Fe/H] = 0) 
and [K/Fe] = +0.5 for the other metal-deficient models 
([Fe/H] = $-1$, $-2$, and $-3$). Namely, the solar potassium abundance 
of 5.12 (Anders, Grevesse 1989; Grevesse, Sauval 2000) was adopted 
for the normal-metal models, 
while a metallicity-scaled potassium abundance plus 0.5 dex (allowing for 
the characteristics suggested from two observational studies so far; 
cf. section 1) was assigned to the metal-poor models.
The microturbulent velocity (appearing in the line-opacity calculations
along with the abundance) was assumed to be 2~km~s$^{-1}$, to make it 
consistent with the model atmosphere.

In figure 3 are shown the $S_{\rm L}(\tau)/B(\tau)$ (the ratio of 
the line source function to the Planck function, and nearly equal to 
$\simeq b_{2}/b_{1}$,
where $b_{1}$ and $b_{2}$ are the non-LTE departure coefficients for
the lower and upper levels of the K~{\sc i}~7699 transition, respectively) 
and  $l_{0}^{\rm NLTE}(\tau)/l_{0}^{\rm LTE}(\tau)$ (the NLTE-to-LTE 
line-center opacity ratio, and nearly equal to $\simeq b_{1}$) for 
a representative set of model atmospheres.
We can read the following characteristics from this figure:\\
--- In almost all cases, the inequality relations of $S_{\rm L}/B<1$ 
(dilution of line source function) and $l_{0}^{\rm NLTE}/l_{0}^{\rm LTE} >1$ 
(enhanced line-opacity) hold in the important line-forming region, which
means that the non-LTE effect always acts in the direction of strengthening
the K~{\sc i}~7699 line.\\
--- There is a tendency that the non-LTE effect is enhanced
 with a lowering of the gravity, as expected.\\
--- The departure from LTE appears to be larger for higher $T_{\rm eff}$
in the high-metallicity (1$\times$) case, while this trend becomes
ambiguous, or even inverse, in the low-metallicity case.\\
--- Toward a lower metallicity, the extent of the non-LTE departure 
tends to decrease, but the departure appears to penetrate deeper
in the atmosphere, which makes the situation rather complex.\\
--- For a very strong damping-dominated case (i.e., lowest $T_{\rm eff}$
and highest metallicity), the departure from LTE shifts toward the upper
atmosphere and the non-LTE effect becomes comparatively insignificant.

Based on the results of these calculations, we computed an extensive
grid of the theoretical equivalent-widths and the corresponding non-LTE
corrections of the K~{\sc i}~7699 line for each of the model atmospheres, 
which are presented in the Appendix.

\section{Abundance Analysis}

Regarding $T_{\rm eff}$ (effective temperature), $\log g$ (surface 
gravity), [Fe/H] (model metallicity), and $\xi$ (microturbulence), 
we simply adopted the same values as those presented in Chen et al. (2000)
and Gratton and Sneden (1987a), where we assigned the [Fe/H] values
spectroscopically determined by them to the model-metallicity
(i.e., not the same model-metallicity as adopted by them).
The solar $\xi$ value was assumed to be 1.4~km~s$^{-1}$ 
(Y.-Q. Chen, unpublished), 
which was determined in the same way using the Fe~{\sc i} lines
as was done by Chen et al. (2000). 
Although this is appreciably larger than that derived from 
Takeda et al.'s (1996) solar flux spectrum analysis of K~{\sc i} 7699
as the best value (0.8~km~s$^{-1}$; cf. subsubsection 5.1.1 therein),
we preferred to adopt this Fe~{\sc i}-based $\xi$ value in order to
maintain the consistency of the analysis, considering the differential
nature of [K/Fe] in which we are most interested.
These atmospheric parameters are given in table 1.

As for the model atmospheres, Kurucz's (1993) grid of ATLAS9 models 
was used as in the case of non-LTE calculations,
based on which the model of each star was obtained by a three-dimensional 
interpolation with respect to $T_{\rm eff}$, $\log g$, and [Fe/H].
Similarly, the depth-dependent departure coefficients ($b$) of 
neutral-potassium levels computed for the grid of models 
(cf. section 3) were interpolated in terms of $T_{\rm eff}$, $\log g$, 
and [Fe/H], in order to evaluate the $S_{\rm L}(\tau)/B(\tau)$ and 
$l_{0}^{\rm NLTE}(\tau)/l_{0}^{\rm LTE}(\tau)$ ratios for each star.

We used the WIDTH9 program written by R. L. Kurucz for determining
the potassium abundance from the K~{\sc i}~7699 equivalent width,
which had been modified to incorporate the non-LTE departure
in the line source function as well as in the line opacity.
The adopted line data for the K~{\sc i} line at 7698.98 $\rm\AA$ are 
essentially the same as those used or determined by Takeda et al. (1996):
$\log gf = -0.17$ (Wiese et al. 1969; NIST database) for the oscillator 
strength, $\Gamma_{\rm rad} = 0.38 \times 10^{8} \; {\rm s}^{-1}$ 
(Wiese et al. 1969; NIST database)
for the radiation damping constant, and $\Delta\log C_{6} = +1.0$
(Takeda et al. 1996) for the correction applied to the classical
Uns\"old's (1955) formula for the van der Waals effect damping constant
(i.e., $\log\Gamma_{6} = \log\Gamma_{6}^{\rm classical}
+ 0.4 \Delta\log C_{6}$).
Regarding the quadratic Stark effect damping (which is insignificant 
in late-type stars), we followed the default treatment of the WIDTH9 
program (cf. Leushin, Topil'skaya 1987). 

The resulting non-LTE abundance ($\log\epsilon_{\rm K}^{\rm NLTE}$) and
the non-LTE abundance correction ($\equiv \log\epsilon_{\rm K}^{\rm NLTE} - 
\log\epsilon_{\rm K}^{\rm LTE}$) are presented in table 1, where
the [K/Fe] values (K-to-Fe logarithmic abundance ratio relative to the Sun),
defined as $\log\epsilon_{\rm K}^{\rm NLTE}({\rm star}) - 
\log\epsilon_{\rm K}^{\rm NLTE}({\rm Sun}) - [{\rm Fe}/{\rm H}]$,
are also given.

Although our reference solar potassium abundance (4.85) is apparently
smaller than the standard solar value of 5.12 (Anders, Grevesse 1989;
Grevesse, Sauval 2000) , this is 
due to our adopted $\xi$ value of 1.4~km~s$^{-1}$, mentioned above,
as well as the use of the ATLAS9 solar model.
If we use Holweger and M\"uller's (1974) model with a $\xi$ value
of 0.8--1.0~km~s$^{-1}$, we would obtain a value of $\sim 5.1$
(cf. table 4 in Takeda et al. 1996).

\section{Error Estimations}

Some discussion may be due concerning the uncertainty in the resulting
abundances. Thanks to the simple ionization nature of potassium atoms,
where almost all of the potassium atoms are in the ground state of 
the first-ionized stage, and only a small fraction of them remain neutral, 
the number population of the ground level of the neutral atoms ($n_{1}$, 
which is proportional to the line opacity, $l$, of the K~{\sc i} 7699 line) is 
expressed as 
$n_{1} \propto \epsilon \theta^{3/2} n_{\rm e} 10^{\chi_{\rm I} \theta}$
according to Saha's equation ($\epsilon$ is the potassium abundance, 
$\theta \equiv 5040/T$, $\chi_{\rm I}$ is 
the ionization potential of 4.34~eV, and $n_{\rm e}$ is the electron density).
Then, the dependence of the line-opacity ($l$) upon $\theta_{\rm eff} 
(\equiv 5040/T_{\rm eff}$) and $g$ may be written as
$l \propto \epsilon \theta_{\rm eff}^{3/2} g^{1/3} 10^{\chi_{\rm I}, 
\theta_{\rm eff}}$
where we put $\theta \sim \theta_{\rm eff}$ and used the approximation 
that the atmospheric density (pressure) 
roughly scales as $\propto g^{1/3}$ (see, e.g., Gray 1992).
Consequently, the abundance ($\epsilon$) resulting from a given
equivalent width is dependent upon 
$\theta_{\rm eff}$ and $g$ as $\epsilon \propto 
\theta_{\rm eff}^{-3/2}10^{-\chi_{\rm I}\theta_{\rm eff}}g^{-1/3}$,
which suggests that changes of $\Delta T_{\rm eff} = \pm 200$~K
and $\Delta\log g= \pm 0.3$ produce variations of $\sim \pm 0.15$
and $\sim \pm 0.10$ (for the case of $T_{\rm eff} = 6000$~K and 
$\log g = 4$ atmosphere), respectively.
In table 1 are given the actually computed changes corresponding to
these perturbations of $T_{\rm eff}$ and $\log g$, which are
to an order of magnitude consistent with such a rough analytical
estimation.\footnote{That the extent of $\Delta_{g}^{\pm}$ is somewhat 
smaller than the analytical prediction may presumably be due to the effect
of the continuum opacity which is also affected by a density variation
through the population of H$^{-}$ ions.}

Another important factor of uncertainty is the microturbulent velocity.
Although most of the $\xi$ values adopted in this study have been 
reasonably established by Chen et al. (2000) and Gratton and Sneden (1987a)
in a conventional way using the Fe~{\sc i} lines, 
there is no guarantee that such values are safely applicable
to an analysis of the strong K~{\sc i} line, because an adequate value
of this parameter differs from line to line, reflecting the possible
depth-dependence of the atmospheric velocity field
(cf. subsubsection 5.1.1 in Takeda et al. 1996).
As a matter of fact, by comparing the adopted values of the microturbulence
with the estimated values based on the empirical formula proposed by
Edvardsson et al. (1993), which is applicable to dwarf stars,
we found that Chen et al.'s (2000) $\xi$ values for BAO samples
are slightly larger [by 0.2 ($\pm 0.2$) km~s$^{-1}$ on the average],
while the $\xi$ values adopted by Gratton and Sneden (1987a)
for their high-gravity samples turn out to be somewhat smaller
[by $-0.3$ ($\pm 0.5$) km~s$^{-1}$ on the average], compared to 
the formula values.
Hence, it should be kept in mind that rather significant ambiguities 
are involved
in the $\xi$ values given in table 1. We obtained abundance variations
corresponding to changes of $\pm 0.5$~km~s$^{-1}$ (tentatively assigned
uncertainty), which amount to $\sim$~0.1--0.2 dex, as given in table 1.

We also evaluated the errors caused by ambiguities in the damping parameter, 
for which van der Waals effect damping ($\Gamma_{6}$) is most important 
in the present case. Since Takeda et al. (1996) concluded the most 
adequate $\Delta\log C_{6}$ value to be $+1.0 \; (\pm 0.4)$ (cf. subsubsection 
5.1.3 therein), we computed the abundance variations corresponding to 
this uncertainty range, which are also presented in table 1. As can be 
seen from this table, they are typically $\ltsim 0.1$~dex and may be 
comparatively less significant.

Based on what has been described above, it would be reasonable to state
that the potassium abundances we have obtained are inevitably subject to 
rather large ambiguities amounting to $\ltsim$~0.2--0.3~dex.

\section{Discussion and Conclusion}

The finally resulting [K/Fe] vs. [Fe/H] relation and $\Delta\log\epsilon$
(non-LTE correction) vs. [Fe/H] relation are shown in figures 4a and b, 
respectively. 

In figure 4a, we can observe a trend that [K/Fe] gradually increases 
with a decrease in [Fe/H] for disk stars, from [K/Fe]~$\sim 0$ 
(at [Fe/H]~$\sim 0$) to [K/Fe]~$\sim$~0.3--0.4 (at [Fe/H]~$\sim -1$), 
and appears to be nearly flat (or showing a weak further increase?) 
at the halo star region 
([Fe/H]~$\ltsim -1$), which is similar to that of the $\alpha$-process
elements (e.g., Mg, Si, Ca, etc.) as is well known.
Especially, a systematic tight correlation exhibited by BAO sample stars 
(filled circles) is quite impressive.
We also note that no clear difference exists between giants and dwarfs, 
as recognized from Gratton and Sneden's (1987) samples (open symbols).
This tendency is almost the same as the results which Gratton and Sneden 
(1987a) and Chen et al. (2000) obtained in their LTE analyses.

As can be seen from figure 4b, the potassium abundances suffer considerably
large non-LTE corrections amounting to 0.2--0.7~dex ($\sim 0.5$~dex on 
the average), which suggests that LTE is by no means an adequate 
assumption for determining the potassium abundance from the strong 
resonance K~{\sc i} line. However, since the extents of these corrections 
do not show any appreciable [Fe/H]-dependence (except for the 
high-metallicity region of [Fe/H]~$\gtsim 0$), the resultant non-LTE 
[K/Fe] vs. [Fe/H] diagram turned out to be almost similar to the LTE case.
This is the reason why our new non-LTE analysis has confirmed the previous
results.

In summary, the conclusions are as follows, based on our non-LTE analysis
of the K~{\sc i}~7699 line data of metal-poor stars:\\
---(1) This resonance K~{\sc i} line suffers a considerable non-LTE effect
(0.2--0.7~dex in terms of the abundance correction) which should be
taken into consideration. However, since the non-LTE correction does not
show any strong systematic [Fe/H]-dependence, its influence is not 
very significant as far as [K/Fe] (differential abundance ratio 
relative to the Sun) is concerned.\\
---(2) We confirmed the previously reported results that [K/Fe] gradually 
increases with a decrease in [Fe/H] for disk stars by $\sim$~0.3--0.4~dex
over the range $-1 \ltsim$~[K/Fe]~$\ltsim 0$, and appears 
to be nearly flat (or slightly increasing further) at the halo star 
region of [K/Fe]~$\ltsim -1$.\\
---(3) Current standard galactic chemical evolution calculations,
such as that of Timmes et al. (1995) using WW95 yields (as they are) and 
Salpeter IMF, do not appear to satisfactorily reproduce this behavior of 
supersolar [K/Fe] in metal-poor stars. In order to bring the theory 
into consistency with the observation, one has to invoke (i) some adjustment
of the yields  [e.g., reducing the WW95 Fe yield by a factor of 2 as suggested 
by Timmes et al. (1995), increasing the WW95 K yield as was done by 
Samland et al. (1998)] and/or (ii) the use of more realistic IMF [e.g., 
that of Kroupa et al. (1993) adopted by Goswami and Prantzos (2000)].

We thank Dr. N. Prantzos and an anonymous referee for pointing out
our misunderstanding in the first version of this paper
concerning the galactic chemical evolution calculations for K.
This work was done within the framework of the China--Japan collaboration
project, ``Galactic Chemical Evolution through Spectrosopic Analyses
of Metal-Deficient Stars'' supported by the Japan Society for the Promotion 
of Sience (JSPS) and the National Science Foundation of China (NSFC).

\appendix
\section{Non-LTE Corrections for a Grid of Models}

For the reader's convenience, we present here an extensive grid of
non-LTE abundance corrections for model atmospheres of various parameters,
which have been computed as follows.

For an appropriately assigned potassium abundance ($A^{\rm a}$) and
microturbulence ($\xi^{\rm a}$), we first 
calculated the non-LTE equivalent width ($W^{\rm NLTE}$) of the line 
by using the computed non-LTE departure coefficients ($b$) for each model 
atmosphere. Next, the LTE ($A^{\rm L}$) and NLTE ($A^{\rm N}$) abundances
were computed from this $W^{\rm NLTE}$
while regarding it as if being a given observed equivalent width.
We can then obtain the non-LTE abundance correction, $\Delta$, which is 
defined in terms of these two abundances as 
$\Delta \equiv A^{\rm N} - A^{\rm L}$.

Strictly speaking, the departure coefficients [$b(\tau)$] for 
a model atmosphere correspond to the potassium abundance and 
the microturbulence of $\log\epsilon_{\rm K}^{\rm input}$
and 2~km~s$^{-1}$ adopted in the non-LTE calculations (cf. section 3). 
Nevertheless, considering the fact that the departure coefficients
(i.e., {\it ratios} of NLTE to LTE number populations) are
(unlike the population itself) not much sensitive to small changes in 
atmospheric parameters,\footnote{For example, a test analysis on
the t60g40m1 model (with $W_{\lambda}^{7699}$ = 158~m$\rm\AA$
and $\xi$ = 2 km~s$^{-1}$) showed that the use of two kinds of $b$ values
computed with two different $\log\epsilon_{\rm K}^{\rm input}$
corresponding to [K/Fe] = 0.0 and +0.5 leads to almost the same
non-LTE abundances (i.e., the differences amounting to only $\sim 0.03$ dex).}
 we also applied such computed $b$ values to 
evaluating $\Delta$ for slightly different $A^{\rm a}$ and $\xi^{\rm a}$ 
from those fiducial values assumed in the statistical equilibrium 
calculations.

Hence, we evaluated $\Delta$ for three $A^{\rm a}$ values
($\log\epsilon_{\rm K}^{\rm input}$ and $\pm 0.3$ dex perturbation)
as well as three $\xi$ values (2~km~s$^{-1}$ and $\pm 1$~km~s$^{-1}$ 
perturbation) for a model atmosphere using the same departure coefficients.
We used the WIDTH9 program with the same line data as in section 4 
for calculating the equivalent width for a given abundance, 
or inversely evaluating the abundance for an assigned equivalent width.
We give only the $\xi$ = 2~km~s$^{-1}$ results for the K~{\sc i}~7699
line in table 2.
The complete data tables ($\xi$ = 1, 2, and 3~km~s$^{-1}$), which 
have been computed not only for the K~{\sc i}~7699 line but also for
the K~{\sc i}~7665 line ($\lambda = 7664.87 \rm\AA$, $\log gf= +0.13$), 
along with a small Fortran program for using/interpolating these tables,
are electronically available from the following anonymous ftp site:\\
$ftp://www.ioa.s.u-tokyo.ac.jp/Users/takeda/potassium\_nonlte/$ \\
(IP address: 133.11.160.242).

\clearpage
\setcounter{table}{0}
\begin{table}[h]
\scriptsize
\caption{Adopted Data of the Program Stars and the Results of the Analysis.}
\begin{center}
\begin{tabular}{
c
c@{ }c@{ }r@{ }c
r@{ }r@{ }c@{ }r
c@{ }c@{ }c@{ }c@{ }c@{ }c@{ }c@{ }c}\hline\hline
Star & $T_{\rm eff}$ & $\log g$ & [Fe/H] & $\xi$ & $W_{\lambda}$ & $\Delta\log\epsilon$ & $\log\epsilon$ & [K/Fe] & $\Delta_{\xi}^{-}$ & $\Delta_{\xi}^{+}$  & $\Delta_{\rm C}^{-}$ & $\Delta_{\rm C}^{+}$ & $\Delta_{T}^{-}$ & $\Delta_{T}^{+}$ & $\Delta_{g}^{-}$ & $\Delta_{g}^{+}$\\
(HD)  & (K) & (cm~s$^{-2}$) &  & (km~s$^{-1}$) & (m$\rm\AA$) &  &  &  &  &  &  &  \\
\hline
\multicolumn{13}{c}{[Analysis of the Solar Flux Equivalent Width]}\\
Sun & 5780 & 4.44  & 0.00  & 1.4  & 169.9  & $-$0.38  & 4.85  & 0.00  &+0.12  & $-$0.13  &+0.07  & $-$0.08          & $-$0.16  & +0.15  & +0.08  & $-$0.09  \\
\hline
\multicolumn{13}{c}{[Analysis of the Remeasured Data of BAO Spectra]}\\
34411  & 5773 & 4.02  &+0.01  & 1.7  & 162.6  & $-$0.48  & 4.76  & $-$0.10  &+0.16  & $-$0.15  &+0.03  & $-$0.05    & $-$0.16  & +0.15  & +0.05  & $-$0.05  \\
19373  & 5867 & 4.01  &+0.03  & 1.8  & 165.9  & $-$0.51  & 4.82  & $-$0.06  &+0.17  & $-$0.15  &+0.03  & $-$0.04    & $-$0.16  & +0.15  & +0.04  & $-$0.05  \\
10307  & 5776 & 4.13  & $-$0.05  & 1.8  & 167.5  & $-$0.48  & 4.74  & $-$0.06  &+0.16  & $-$0.14  &+0.04  & $-$0.05 & $-$0.16  & +0.15  & +0.05  & $-$0.06  \\
68146  & 6227 & 4.16  & $-$0.09  & 2.1  & 144.9  & $-$0.57  & 4.73  & $-$0.03  &+0.13  & $-$0.11  &+0.01  & $-$0.03 & $-$0.14  & +0.14  & +0.03  & $-$0.03  \\
22484  & 5915 & 4.03  & $-$0.13  & 2.0  & 163.2  & $-$0.57  & 4.72  &  0.00  &+0.15  & $-$0.13  &+0.02  & $-$0.04   & $-$0.16  & +0.15  & +0.04  & $-$0.04  \\
39587  & 5805 & 4.29  & $-$0.18  & 2.2  & 168.6  & $-$0.49  & 4.60  & $-$0.07  &+0.14  & $-$0.11  &+0.04  & $-$0.04 & $-$0.16  & +0.15  & +0.04  & $-$0.05  \\
142860 & 6227 & 4.18  & $-$0.22  & 2.2  & 132.8  & $-$0.56  & 4.57  & $-$0.06  &+0.12  & $-$0.08  &+0.02  & $-$0.02 & $-$0.14  & +0.13  & +0.02  & $-$0.03  \\
69897  & 6243 & 4.28  & $-$0.28  & 2.0  & 136.9  & $-$0.58  & 4.64  & +0.07  &+0.12  & $-$0.09  &+0.03  & $-$0.02   & $-$0.14  & +0.13  & +0.03  & $-$0.03  \\
55575  & 5802 & 4.36  & $-$0.36  & 1.6  & 141.7  & $-$0.48  & 4.47  & $-$0.02  &+0.12  & $-$0.11  &+0.04  & $-$0.05 & $-$0.16  & +0.16  & +0.05  & $-$0.06  \\
142373 & 5920 & 4.27  & $-$0.39  & 1.5  & 137.9  & $-$0.54  & 4.54  & +0.08  &+0.12  & $-$0.11  &+0.04  & $-$0.05   & $-$0.16  & +0.15  & +0.05  & $-$0.06  \\
101676 & 6102 & 4.09  & $-$0.47  & 2.0  & 131.6  & $-$0.61  & 4.49  & +0.11  &+0.12  & $-$0.09  &+0.02  & $-$0.02   & $-$0.14  & +0.14  & +0.03  & $-$0.03  \\
76349  & 6004 & 4.21  & $-$0.49  & 2.1  & 132.0  & $-$0.56  & 4.40  & +0.04  &+0.10  & $-$0.08  &+0.02  & $-$0.03   & $-$0.15  & +0.14  & +0.03  & $-$0.03  \\
118244 & 6234 & 4.13  & $-$0.55  & 2.3  & 126.0  & $-$0.60  & 4.46  &+0.16  &+0.10  & $-$0.07  &+0.01  & $-$0.02    & $-$0.13  & +0.12  & +0.02  & $-$0.02  \\
109303 & 5905 & 4.10  & $-$0.61  & 1.7  & 122.2  & $-$0.58  & 4.32  & +0.08  &+0.11  & $-$0.09  &+0.03  & $-$0.03   & $-$0.15  & +0.14  & +0.03  & $-$0.04  \\
41640  & 6004 & 4.37  & $-$0.62  & 2.0  & 114.8  & $-$0.51  & 4.25  & +0.02  &+0.08  & $-$0.07  &+0.02  & $-$0.03   & $-$0.15  & +0.14  & +0.03  & $-$0.03  \\
62301  & 5837 & 4.23  & $-$0.67  & 1.7  & 126.9  & $-$0.54  & 4.30  & +0.12  &+0.11  & $-$0.09  &+0.04  & $-$0.04   & $-$0.16  & +0.15  & +0.04  & $-$0.05  \\
49732  & 6260 & 4.15  & $-$0.70  & 1.9  & 104.5  & $-$0.58  & 4.34  &+0.19  &+0.09  & $-$0.07  &+0.01  & $-$0.02    & $-$0.13  & +0.12  & +0.02  & $-$0.02  \\
106516 & 6135 & 4.34  & $-$0.71  & 1.5  & 112.6  & $-$0.57  & 4.39  &+0.25  &+0.10  & $-$0.08  &+0.03  & $-$0.03    & $-$0.14  & +0.13  & +0.04  & $-$0.04  \\
208906 & 5929 & 4.39  & $-$0.73  & 1.5  & 112.5  & $-$0.52  & 4.25  & +0.13  &+0.08  & $-$0.08  &+0.03  & $-$0.04   & $-$0.15  & +0.14  & +0.04  & $-$0.05  \\
60319  & 5867 & 4.24  & $-$0.85  & 1.6  & 117.2  & $-$0.55  & 4.24  &+0.24  &+0.09  & $-$0.08  &+0.03  & $-$0.04    & $-$0.16  & +0.14  & +0.04  & $-$0.05  \\
201891 & 5827 & 4.43  & $-$1.04  & 1.6  & 101.4  & $-$0.47  & 4.04  &+0.23  &+0.07  & $-$0.06  &+0.03  & $-$0.04    & $-$0.16  & +0.14  & +0.04  & $-$0.04  \\
\hline                                             
\multicolumn{13}{c}{[Reanalysis of Gratton and Sneden's (1987b) Data]}\\
4614   & 5670 & 4.30  & $-$0.33  & 1.0  & 122.0  & $-$0.43  & 4.35  & $-$0.17  &+0.09  & $-$0.11  &+0.05  & $-$0.06 & $-$0.17  & +0.16  & +0.06  & $-$0.07  \\
6582   & 5250 & 4.50  & $-$0.75  & 1.0  & 145.0  & $-$0.34  & 4.07  & $-$0.03  &+0.06  & $-$0.08  &+0.09  & $-$0.09 & $-$0.21  & +0.19  & +0.09  & $-$0.10  \\
6755   & 5260 & 3.00  & $-$1.67  & 2.7  & 93.9  & $-$0.41  & 3.53  &+0.35  &+0.05  & $-$0.04  &+0.00  & $-$0.01     & $-$0.18  & +0.15  & +0.02  & $-$0.02  \\
33256  & 6140 & 3.91  & $-$0.58  & 1.2  & 121.0  & $-$0.69  & 4.61  &+0.34  &+0.12  & $-$0.13  &+0.02  & $-$0.04    & $-$0.14  & +0.13  & +0.04  & $-$0.05  \\
34411  & 5840 & 4.10  &+0.16  & 1.0  & 156.0  & $-$0.42  & 4.99  & $-$0.02  &+0.12  & $-$0.17  &+0.06  & $-$0.08    & $-$0.16  & +0.15  & +0.07  & $-$0.08  \\
58551  & 6000 & 4.50  & $-$0.78  & 1.0  & 112.0  & $-$0.52  & 4.35  &+0.28  &+0.07  & $-$0.08  &+0.05  & $-$0.06    & $-$0.15  & +0.14  & +0.06  & $-$0.07  \\
59984  & 5840 & 4.01  & $-$0.93  & 1.0  & 121.4  & $-$0.60  & 4.41  &+0.49  &+0.09  & $-$0.11  &+0.04  & $-$0.06    & $-$0.16  & +0.15  & +0.06  & $-$0.07  \\
63077  & 5660 & 4.11  & $-$0.94  & 0.8  & 136.3  & $-$0.51  & 4.43  &+0.52  &+0.07  & $-$0.10  &+0.08  & $-$0.08    & $-$0.18  & +0.17  & +0.09  & $-$0.09  \\
76932  & 5630 & 3.88  & $-$1.01  & 0.8  & 102.3  & $-$0.55  & 4.10  &+0.26  &+0.07  & $-$0.09  &+0.05  & $-$0.05    & $-$0.17  & +0.16  & +0.06  & $-$0.06  \\
102870 & 6080 & 4.11  &+0.06  & 1.2  & 155.6  & $-$0.51  & 5.04  & +0.13  &+0.15  & $-$0.16  &+0.05  & $-$0.06      & $-$0.15  & +0.14  & +0.06  & $-$0.07  \\
103095 & 5000 & 4.50  & $-$1.15  & 1.0  & 118.0  & $-$0.30  & 3.55  & $-$0.15  &+0.04  & $-$0.06  &+0.07  & $-$0.08 & $-$0.23  & +0.21  & +0.08  & $-$0.08  \\
107328 & 4560 & 2.70  & $-$0.31  & 1.7  & 225.7  & $-$0.27  & 4.50  & $-$0.04  &+0.24  & $-$0.26  &+0.06  & $-$0.07 & $-$0.27  & +0.25  & +0.07  & $-$0.08  \\
122563 & 4640 & 1.30  & $-$2.38  & 2.0  & 37.1  & $-$0.26  & 2.55  & +0.08  &+0.03  & $-$0.02  &+0.00  &+0.00       & $-$0.21  & +0.16  & +0.04  & $-$0.04  \\
122956 & 4660 & 1.75  & $-$1.93  & 1.7  & 120.8  & $-$0.60  & 3.49  &+0.57  &+0.19  & $-$0.13  &+0.01  & $-$0.01    & $-$0.26  & +0.22  & +0.07  & $-$0.06  \\
128620 & 5750 & 4.38  &+0.11  & 1.2  & 174.1  & $-$0.35  & 4.96  & 0.00  &+0.11  & $-$0.14  &+0.08  & $-$0.08       & $-$0.16  & +0.16  & +0.09  & $-$0.10  \\
128621 & 5250 & 4.50  &+0.08  & 1.0  & 250.1  & $-$0.18  & 5.03  & +0.10  &+0.06  & $-$0.08  &+0.13  & $-$0.13      & $-$0.22  & +0.20  & +0.14  & $-$0.14  \\
134169 & 5750 & 4.50  & $-$1.02  & 1.0  & 106.0  & $-$0.46  & 4.09  &+0.26  &+0.06  & $-$0.06  &+0.06  & $-$0.05    & $-$0.17  & +0.16  & +0.06  & $-$0.07  \\
140573 & 4600 & 2.75  &+0.36  & 1.5  & 254.9  & $-$0.13  & 5.07  & $-$0.14  &+0.20  & $-$0.28  &+0.08  & $-$0.09    & $-$0.26  & +0.25  & +0.09  & $-$0.10  \\
142373 & 5760 & 3.90  & $-$0.36  & 1.0  & 121.0  & $-$0.55  & 4.45  & $-$0.04  &+0.11  & $-$0.13  &+0.04  & $-$0.05 & $-$0.16  & +0.16  & +0.05  & $-$0.06  \\
161096 & 4600 & 2.70  &+0.24  & 1.4  & 259.8  & $-$0.12  & 5.13  & +0.04  &+0.17  & $-$0.25  &+0.09  & $-$0.10      & $-$0.27  & +0.26  & +0.10  & $-$0.11  \\
165195 & 4500 & 1.20  & $-$2.21  & 2.0  & 115.0  & $-$0.49  & 3.27  &+0.63  &+0.16  & $-$0.11  &+0.01  &+0.00       & $-$0.29  & +0.21  & +0.08  & $-$0.08  \\
175305 & 5160 & 3.00  & $-$1.53  & 2.0  & 110.0  & $-$0.52  & 3.68  &+0.36  &+0.11  & $-$0.08  &+0.02  & $-$0.01    & $-$0.20  & +0.17  & +0.03  & $-$0.03  \\
221170 & 4500 & 1.30  & $-$1.96  & 1.4  & 134.0  & $-$0.63  & 3.62  &+0.73  &+0.27  & $-$0.21  &+0.01  & $-$0.02    & $-$0.30  & +0.25  & +0.09  & $-$0.09  \\
224930 & 5250 & 4.50  & $-$0.84  & 1.0  & 177.0  & $-$0.34  & 4.26  &+0.25  &+0.06  & $-$0.07  &+0.11  & $-$0.10    & $-$0.21  & +0.20  & +0.12  & $-$0.12  \\
\hline
\end{tabular}
\end{center}
Notes. The atmospheric parameters ($T_{\rm eff}$, $\log g$, [Fe/H], and $\xi$) 
were taken from Chen et al. (2000), 
and Gratton and Sneden (1987a), for each data group, respectively. 
Given in the 7th and 8th columns are the non-LTE correction 
($\equiv \log\epsilon_{\rm K}^{\rm NLTE} - \log\epsilon_{\rm K}^{\rm LTE}$)
and the non-LTE abundance ($\log\epsilon_{\rm K}^{\rm NLTE}$; in the usual 
normalization of $\log\epsilon_{\rm H} = 12.00$), respectively.
The K to Fe logarithmic abundance ratio, [K/Fe], in the 9th column
is defined as $\log\epsilon_{\rm K}^{\rm NLTE}({\rm star}) - 
\log\epsilon_{\rm K}^{\rm NLTE}({\rm Sun}) - [{\rm Fe}/{\rm H}]$.
The abundance variations caused by changing the adopted parameters
are given in 10th through 17th columns: changing the $\xi$ value 
by $-$0.5~km~s$^{-1}$ ($\Delta_{\xi}^{-}$) and +0.5~km~s$^{-1}$
($\Delta_{\xi}^{+}$), changing the $\Delta\log C_{6}$ value 
(with respect to the fiducial value of +1.0) by $-0.4$ ($\Delta_{\rm C}^{-}$) 
and +0.4 ($\Delta_{\rm C}^{+}$), changing the $T_{\rm eff}$ value 
by $-200$~K ($\Delta_{T}^{-}$) and +200~K ($\Delta_{T}^{+}$),
and changing the $\log g$ value by $-0.3$ ($\Delta_{g}^{-}$) and 
+0.3 ($\Delta_{g}^{+}$).
\end{table}

\clearpage
\setcounter{table}{1}
\begin{table}[h]
\caption{NLTE abundance correction for K I 7698.98
 ($\xi$ = 2 ${\rm km}\, {\rm s}^{-1}$).}
\scriptsize
\begin{center}
\begin{tabular}
{c
r@{ }r@{ }r@{ }r@{ }r@{ }r
r@{ }r@{ }r@{ }r@{ }r@{ }r
r@{ }r@{ }r@{ }r@{ }r@{ }r}\hline \hline
Code
& $A_{1}^{\rm a}$ & ($W_{1}^{\rm LTE}$) & $W_{1}^{\rm NLTE}$ 
& $A_{1}^{\rm N}$ & $A_{1}^{\rm L}$ & $\Delta_{1}$
& $A_{2}^{\rm a}$ & ($W_{2}^{\rm LTE}$) & $W_{2}^{\rm NLTE}$
& $A_{2}^{\rm N}$ & $A_{2}^{\rm L}$ & $\Delta_{2}$
& $A_{3}^{\rm a}$ & ($W_{3}^{\rm LTE}$) & $W_{3}^{\rm NLTE}$
& $A_{3}^{\rm N}$ & $A_{3}^{\rm L}$ & $\Delta_{3}$  \\
\hline
t65g50m0
& 4.820 & ( 97.72) &  138.04
& 4.819 &  5.236
&  $-$0.417
& 5.120 & (125.89) &  177.83
& 5.120 &  5.562
&  $-$0.442
& 5.420 & (158.49) &  223.87
& 5.425 &  5.847
&  $-$0.422 \\
t65g40m0
& 4.820 & ( 87.10) &  131.83
& 4.812 &  5.439
&  $-$0.627
& 5.120 & (107.15) &  165.96
& 5.130 &  5.840
&  $-$0.710
& 5.420 & (131.83) &  199.53
& 5.439 &  6.149
&  $-$0.710 \\
t65g30m0
& 4.820 & ( 81.28) &  131.83
& 4.828 &  5.590
&  $-$0.762
& 5.120 & (102.33) &  162.18
& 5.130 &  6.043
&  $-$0.913
& 5.420 & (120.23) &  190.55
& 5.427 &  6.409
&  $-$0.982 \\
t65g20m0
& 4.820 & ( 83.18) &  134.90
& 4.830 &  5.651
&  $-$0.821
& 5.120 & (102.33) &  162.18
& 5.105 &  6.094
&  $-$0.989
& 5.420 & (120.23) &  190.55
& 5.413 &  6.519
&  $-$1.106 \\
t65g10m0
& 4.820 & ( 75.86) &  125.89
& 4.816 &  5.613
&  $-$0.797
& 5.120 & ( 95.50) &  154.88
& 5.107 &  6.096
&  $-$0.989
& 5.420 & (114.82) &  181.97
& 5.398 &  6.537
&  $-$1.139 \\
t60g50m0
& 4.820 & (138.04) &  181.97
& 4.827 &  5.171
&  $-$0.344
& 5.120 & (173.78) &  229.09
& 5.117 &  5.449
&  $-$0.332
& 5.420 & (223.87) &  288.40
& 5.410 &  5.708
&  $-$0.298 \\
t60g40m0
& 4.820 & (117.49) &  165.96
& 4.827 &  5.390
&  $-$0.563
& 5.120 & (141.25) &  199.53
& 5.117 &  5.685
&  $-$0.568
& 5.420 & (169.82) &  239.88
& 5.437 &  5.953
&  $-$0.516 \\
t60g30m0
& 4.820 & (107.15) &  162.18
& 4.836 &  5.597
&  $-$0.761
& 5.120 & (128.82) &  190.55
& 5.120 &  5.935
&  $-$0.815
& 5.420 & (147.91) &  218.78
& 5.410 &  6.209
&  $-$0.799 \\
t60g20m0
& 4.820 & (102.33) &  158.49
& 4.826 &  5.681
&  $-$0.855
& 5.120 & (123.03) &  186.21
& 5.118 &  6.092
&  $-$0.974
& 5.420 & (141.25) &  213.80
& 5.428 &  6.447
&  $-$1.019 \\
t60g10m0
& 4.820 & (102.33) &  158.49
& 4.823 &  5.701
&  $-$0.878
& 5.120 & (123.03) &  186.21
& 5.120 &  6.136
&  $-$1.016
& 5.420 & (141.25) &  213.80
& 5.439 &  6.531
&  $-$1.092 \\
t55g50m0
& 4.820 & (199.53) &  245.47
& 4.832 &  5.074
&  $-$0.242
& 5.120 & (257.04) &  309.03
& 5.110 &  5.328
&  $-$0.218
& 5.420 & (338.84) &  407.38
& 5.429 &  5.618
&  $-$0.189 \\
t55g40m0
& 4.820 & (154.88) &  204.17
& 4.832 &  5.254
&  $-$0.422
& 5.120 & (186.21) &  245.47
& 5.135 &  5.520
&  $-$0.385
& 5.420 & (229.09) &  295.12
& 5.432 &  5.763
&  $-$0.331 \\
t55g30m0
& 4.820 & (141.25) &  190.55
& 4.816 &  5.435
&  $-$0.619
& 5.120 & (162.18) &  223.87
& 5.145 &  5.749
&  $-$0.604
& 5.420 & (190.55) &  251.19
& 5.400 &  5.955
&  $-$0.555 \\
t55g20m0
& 4.820 & (134.90) &  186.21
& 4.807 &  5.560
&  $-$0.753
& 5.120 & (154.88) &  213.80
& 5.111 &  5.900
&  $-$0.789
& 5.420 & (177.83) &  239.88
& 5.404 &  6.168
&  $-$0.764 \\
t55g10m0
& 4.820 & (128.82) &  186.21
& 4.844 &  5.675
&  $-$0.831
& 5.120 & (147.91) &  208.93
& 5.100 &  5.997
&  $-$0.897
& 5.420 & (169.82) &  234.42
& 5.406 &  6.315
&  $-$0.909 \\
t50g50m0
& 4.820 & (323.59) &  371.54
& 4.819 &  4.965
&  $-$0.146
& 5.120 & (436.52) &  489.78
& 5.109 &  5.238
&  $-$0.129
& 5.420 & (588.84) &  676.08
& 5.425 &  5.541
&  $-$0.116 \\
t50g40m0
& 4.820 & (218.78) &  263.03
& 4.823 &  5.070
&  $-$0.247
& 5.120 & (275.42) &  323.59
& 5.127 &  5.332
&  $-$0.205
& 5.420 & (346.74) &  407.38
& 5.429 &  5.596
&  $-$0.167 \\
t50g30m0
& 4.820 & (186.21) &  229.09
& 4.826 &  5.217
&  $-$0.391
& 5.120 & (218.78) &  263.03
& 5.119 &  5.467
&  $-$0.348
& 5.420 & (257.04) &  302.00
& 5.401 &  5.684
&  $-$0.283 \\
t50g20m0
& 4.820 & (177.83) &  218.78
& 4.815 &  5.326
&  $-$0.511
& 5.120 & (199.53) &  245.47
& 5.111 &  5.598
&  $-$0.487
& 5.420 & (229.09) &  275.42
& 5.421 &  5.848
&  $-$0.427 \\
t50g10m0
& 4.820 & (173.78) &  218.78
& 4.850 &  5.422
&  $-$0.572
& 5.120 & (194.98) &  239.88
& 5.105 &  5.668
&  $-$0.563
& 5.420 & (218.78) &  263.03
& 5.384 &  5.914
&  $-$0.530 \\
t45g50m0
& 4.820 & (630.96) &  707.95
& 4.828 &  4.923
&  $-$0.095
& 5.120 & (870.96) &  954.99
& 5.121 &  5.206
&  $-$0.085
& 5.420 & (202.26) &  288.25
& 5.411 &  5.483
&  $-$0.072 \\
t45g40m0
& 4.820 & (380.19) &  416.87
& 4.829 &  4.934
&  $-$0.105
& 5.120 & (501.19) &  537.03
& 5.112 &  5.196
&  $-$0.084
& 5.420 & (676.08) &  724.44
& 5.412 &  5.485
&  $-$0.073 \\
t45g30m0
& 4.820 & (269.15) &  295.12
& 4.811 &  4.961
&  $-$0.150
& 5.120 & (323.59) &  354.81
& 5.121 &  5.234
&  $-$0.113
& 5.420 & (407.38) &  436.52
& 5.411 &  5.492
&  $-$0.081 \\
t45g20m0
& 4.820 & (239.88) &  263.03
& 4.825 &  5.014
&  $-$0.189
& 5.120 & (275.42) &  295.12
& 5.117 &  5.268
&  $-$0.151
& 5.420 & (316.23) &  338.84
& 5.432 &  5.540
&  $-$0.108 \\
t45g10m0
& 4.820 & (239.88) &  251.19
& 4.791 &  4.969
&  $-$0.178
& 5.120 & (263.03) &  281.84
& 5.145 &  5.303
&  $-$0.158
& 5.420 & (295.12) &  309.03
& 5.423 &  5.553
&  $-$0.130 \\
t65g50m1
& 4.320 & ( 57.54) &   93.33
& 4.314 &  4.746
&  $-$0.432
& 4.620 & ( 83.18) &  131.83
& 4.619 &  5.136
&  $-$0.517
& 4.920 & (109.65) &  173.78
& 4.920 &  5.463
&  $-$0.543 \\
t65g40m1
& 4.320 & ( 52.48) &   91.20
& 4.328 &  4.906
&  $-$0.578
& 4.620 & ( 72.44) &  123.03
& 4.626 &  5.372
&  $-$0.746
& 4.920 & ( 91.20) &  154.88
& 4.922 &  5.757
&  $-$0.835 \\
t65g30m1
& 4.320 & ( 48.98) &   87.10
& 4.312 &  4.932
&  $-$0.620
& 4.620 & ( 67.61) &  117.49
& 4.608 &  5.464
&  $-$0.856
& 4.920 & ( 87.10) &  147.91
& 4.914 &  5.966
&  $-$1.052 \\
t65g20m1
& 4.320 & ( 51.29) &   91.20
& 4.314 &  5.000
&  $-$0.686
& 4.620 & ( 69.18) &  123.03
& 4.628 &  5.586
&  $-$0.958
& 4.920 & ( 87.10) &  151.36
& 4.920 &  6.095
&  $-$1.175 \\
t65g10m1
& 4.320 & ( 45.71) &   85.11
& 4.325 &  4.970
&  $-$0.645
& 4.620 & ( 64.57) &  114.82
& 4.619 &  5.509
&  $-$0.890
& 4.920 & ( 83.18) &  144.54
& 4.925 &  6.058
&  $-$1.133 \\
t60g50m1
& 4.320 & ( 87.10) &  138.04
& 4.317 &  4.764
&  $-$0.447
& 4.620 & (120.23) &  186.21
& 4.613 &  5.085
&  $-$0.472
& 4.920 & (158.49) &  245.47
& 4.916 &  5.374
&  $-$0.458 \\
t60g40m1
& 4.320 & ( 74.13) &  123.03
& 4.310 &  4.951
&  $-$0.641
& 4.620 & ( 97.72) &  158.49
& 4.613 &  5.322
&  $-$0.709
& 4.920 & (120.23) &  194.98
& 4.908 &  5.607
&  $-$0.699 \\
t60g30m1
& 4.320 & ( 70.79) &  120.23
& 4.331 &  5.128
&  $-$0.797
& 4.620 & ( 89.13) &  151.36
& 4.632 &  5.585
&  $-$0.953
& 4.920 & (107.15) &  181.97
& 4.940 &  5.937
&  $-$0.997 \\
t60g20m1
& 4.320 & ( 67.61) &  117.49
& 4.332 &  5.172
&  $-$0.840
& 4.620 & ( 85.11) &  147.91
& 4.634 &  5.701
&  $-$1.067
& 4.920 & (102.33) &  173.78
& 4.908 &  6.106
&  $-$1.198 \\
t60g10m1
& 4.320 & ( 69.18) &  117.49
& 4.311 &  5.165
&  $-$0.854
& 4.620 & ( 87.10) &  147.91
& 4.618 &  5.714
&  $-$1.096
& 4.920 & (104.71) &  177.83
& 4.941 &  6.220
&  $-$1.279 \\
t55g50m1
& 4.320 & (141.25) &  208.93
& 4.323 &  4.700
&  $-$0.377
& 4.620 & (194.98) &  275.42
& 4.610 &  4.977
&  $-$0.367
& 4.920 & (257.04) &  371.54
& 4.925 &  5.267
&  $-$0.342 \\
t55g40m1
& 4.320 & (109.65) &  169.82
& 4.335 &  4.895
&  $-$0.560
& 4.620 & (138.04) &  208.93
& 4.611 &  5.158
&  $-$0.547
& 4.920 & (173.78) &  263.03
& 4.932 &  5.428
&  $-$0.496 \\
t55g30m1
& 4.320 & ( 95.50) &  154.88
& 4.335 &  5.117
&  $-$0.782
& 4.620 & (117.49) &  186.21
& 4.627 &  5.432
&  $-$0.805
& 4.920 & (138.04) &  218.78
& 4.922 &  5.681
&  $-$0.759 \\
t55g20m1
& 4.320 & ( 91.20) &  147.91
& 4.318 &  5.248
&  $-$0.930
& 4.620 & (109.65) &  177.83
& 4.626 &  5.663
&  $-$1.037
& 4.920 & (128.82) &  204.17
& 4.910 &  5.957
&  $-$1.047 \\
t55g10m1
& 4.320 & ( 87.10) &  144.54
& 4.319 &  5.292
&  $-$0.973
& 4.620 & (104.71) &  173.78
& 4.628 &  5.778
&  $-$1.150
& 4.920 & (123.03) &  199.53
& 4.920 &  6.151
&  $-$1.231 \\
t50g50m1
& 4.320 & (251.19) &  338.84
& 4.330 &  4.598
&  $-$0.268
& 4.620 & (346.74) &  457.09
& 4.627 &  4.879
&  $-$0.252
& 4.920 & (478.63) &  616.60
& 4.913 &  5.153
&  $-$0.240 \\
t50g40m1
& 4.320 & (169.82) &  234.42
& 4.331 &  4.714
&  $-$0.383
& 4.620 & (218.78) &  295.12
& 4.629 &  4.970
&  $-$0.341
& 4.920 & (281.84) &  371.54
& 4.911 &  5.212
&  $-$0.301 \\
t50g30m1
& 4.320 & (138.04) &  194.98
& 4.313 &  4.887
&  $-$0.574
& 4.620 & (165.96) &  234.42
& 4.633 &  5.157
&  $-$0.524
& 4.920 & (199.53) &  275.42
& 4.918 &  5.372
&  $-$0.454 \\
t50g20m1
& 4.320 & (125.89) &  186.21
& 4.342 &  5.108
&  $-$0.766
& 4.620 & (147.91) &  213.80
& 4.628 &  5.378
&  $-$0.750
& 4.920 & (169.82) &  245.47
& 4.947 &  5.624
&  $-$0.677 \\
t50g10m1
& 4.320 & (120.23) &  177.83
& 4.317 &  5.193
&  $-$0.876
& 4.620 & (138.04) &  204.17
& 4.612 &  5.530
&  $-$0.918
& 4.920 & (158.49) &  229.09
& 4.902 &  5.798
&  $-$0.896 \\
t45g50m1
& 4.320 & (446.68) &  588.84
& 4.326 &  4.565
&  $-$0.239
& 4.620 & (630.96) &  812.83
& 4.629 &  4.859
&  $-$0.230
& 4.920 & (870.96) &   96.48
& 4.916 &  5.132
&  $-$0.216 \\
t45g40m1
& 4.320 & (302.00) &  371.54
& 4.312 &  4.528
&  $-$0.216
& 4.620 & (407.38) &  489.78
& 4.614 &  4.799
&  $-$0.185
& 4.920 & (562.34) &  660.69
& 4.912 &  5.081
&  $-$0.169 \\
t45g30m1
& 4.320 & (213.80) &  263.03
& 4.322 &  4.621
&  $-$0.299
& 4.620 & (263.03) &  316.23
& 4.608 &  4.853
&  $-$0.245
& 4.920 & (338.84) &  398.11
& 4.923 &  5.114
&  $-$0.191 \\
t45g20m1
& 4.320 & (186.21) &  229.09
& 4.312 &  4.728
&  $-$0.416
& 4.620 & (213.80) &  263.03
& 4.622 &  4.975
&  $-$0.353
& 4.920 & (257.04) &  302.00
& 4.917 &  5.193
&  $-$0.276 \\
t45g10m1
& 4.320 & (177.83) &  218.78
& 4.304 &  4.788
&  $-$0.484
& 4.620 & (204.17) &  245.47
& 4.611 &  5.059
&  $-$0.448
& 4.920 & (229.09) &  275.42
& 4.936 &  5.318
&  $-$0.382 \\
\hline
\end{tabular}
\end{center}
\end{table}

\clearpage
\setcounter{table}{1}
\begin{table}[h]
\caption{(Continued.)}
\scriptsize
\begin{center}
\begin{tabular}
{c
r@{ }r@{ }r@{ }r@{ }r@{ }r
r@{ }r@{ }r@{ }r@{ }r@{ }r
r@{ }r@{ }r@{ }r@{ }r@{ }r}\hline \hline
Code
& $A_{1}^{\rm a}$ & ($W_{1}^{\rm LTE}$) & $W_{1}^{\rm NLTE}$ 
& $A_{1}^{\rm N}$ & $A_{1}^{\rm L}$ & $\Delta_{1}$
& $A_{2}^{\rm a}$ & ($W_{2}^{\rm LTE}$) & $W_{2}^{\rm NLTE}$
& $A_{2}^{\rm N}$ & $A_{2}^{\rm L}$ & $\Delta_{2}$
& $A_{3}^{\rm a}$ & ($W_{3}^{\rm LTE}$) & $W_{3}^{\rm NLTE}$
& $A_{3}^{\rm N}$ & $A_{3}^{\rm L}$ & $\Delta_{3}$  \\
\hline
t65g50m2
& 3.320 & ( 10.23) &   16.98
& 3.321 &  3.569
&  $-$0.248
& 3.620 & ( 18.62) &   30.90
& 3.622 &  3.885
&  $-$0.263
& 3.920 & ( 33.11) &   52.48
& 3.921 &  4.217
&  $-$0.296 \\
t65g40m2
& 3.320 & (  9.77) &   16.98
& 3.316 &  3.593
&  $-$0.277
& 3.620 & ( 17.78) &   30.90
& 3.625 &  3.925
&  $-$0.300
& 3.920 & ( 30.90) &   50.12
& 3.913 &  4.256
&  $-$0.343 \\
t65g30m2
& 3.320 & (  8.91) &   16.22
& 3.316 &  3.614
&  $-$0.298
& 3.620 & ( 16.60) &   29.51
& 3.625 &  3.944
&  $-$0.319
& 3.920 & ( 28.18) &   48.98
& 3.927 &  4.293
&  $-$0.366 \\
t65g20m2
& 3.320 & (  9.55) &   17.78
& 3.325 &  3.629
&  $-$0.304
& 3.620 & ( 17.38) &   31.62
& 3.626 &  3.958
&  $-$0.332
& 3.920 & ( 29.51) &   51.29
& 3.920 &  4.313
&  $-$0.393 \\
t65g10m2
& 3.320 & (  8.13) &   15.14
& 3.319 &  3.627
&  $-$0.308
& 3.620 & ( 14.79) &   27.54
& 3.624 &  3.955
&  $-$0.331
& 3.920 & ( 25.70) &   45.71
& 3.919 &  4.297
&  $-$0.378 \\
t60g50m2
& 3.320 & ( 17.38) &   30.20
& 3.326 &  3.600
&  $-$0.274
& 3.620 & ( 31.62) &   52.48
& 3.622 &  3.920
&  $-$0.298
& 3.920 & ( 52.48) &   85.11
& 3.926 &  4.266
&  $-$0.340 \\
t60g40m2
& 3.320 & ( 17.38) &   30.20
& 3.320 &  3.619
&  $-$0.299
& 3.620 & ( 30.20) &   51.29
& 3.624 &  3.966
&  $-$0.342
& 3.920 & ( 47.86) &   77.62
& 3.915 &  4.336
&  $-$0.421 \\
t60g30m2
& 3.320 & ( 16.60) &   29.51
& 3.318 &  3.634
&  $-$0.316
& 3.620 & ( 28.84) &   48.98
& 3.614 &  3.977
&  $-$0.363
& 3.920 & ( 45.71) &   74.13
& 3.912 &  4.376
&  $-$0.464 \\
t60g20m2
& 3.320 & ( 15.49) &   28.18
& 3.317 &  3.646
&  $-$0.329
& 3.620 & ( 26.92) &   47.86
& 3.624 &  4.001
&  $-$0.377
& 3.920 & ( 42.66) &   72.44
& 3.923 &  4.398
&  $-$0.475 \\
t60g10m2
& 3.320 & ( 16.60) &   29.51
& 3.316 &  3.639
&  $-$0.323
& 3.620 & ( 28.84) &   48.98
& 3.613 &  3.992
&  $-$0.379
& 3.920 & ( 44.67) &   74.13
& 3.916 &  4.413
&  $-$0.497 \\
t55g50m2
& 3.320 & ( 31.62) &   56.23
& 3.324 &  3.628
&  $-$0.304
& 3.620 & ( 54.95) &   93.33
& 3.620 &  3.954
&  $-$0.334
& 3.920 & ( 89.13) &  144.54
& 3.923 &  4.296
&  $-$0.373 \\
t55g40m2
& 3.320 & ( 30.90) &   52.48
& 3.314 &  3.655
&  $-$0.341
& 3.620 & ( 50.12) &   83.18
& 3.622 &  4.032
&  $-$0.410
& 3.920 & ( 74.13) &  117.49
& 3.914 &  4.409
&  $-$0.495 \\
t55g30m2
& 3.320 & ( 30.20) &   51.29
& 3.317 &  3.676
&  $-$0.359
& 3.620 & ( 47.86) &   79.43
& 3.628 &  4.094
&  $-$0.466
& 3.920 & ( 67.61) &  109.65
& 3.933 &  4.556
&  $-$0.623 \\
t55g20m2
& 3.320 & ( 28.84) &   50.12
& 3.320 &  3.694
&  $-$0.374
& 3.620 & ( 45.71) &   75.86
& 3.617 &  4.097
&  $-$0.480
& 3.920 & ( 64.57) &  104.71
& 3.923 &  4.592
&  $-$0.669 \\
t55g10m2
& 3.320 & ( 26.92) &   47.86
& 3.315 &  3.696
&  $-$0.381
& 3.620 & ( 43.65) &   74.13
& 3.625 &  4.110
&  $-$0.485
& 3.920 & ( 61.66) &  102.33
& 3.929 &  4.596
&  $-$0.667 \\
t50g50m2
& 3.320 & ( 74.13) &  117.49
& 3.316 &  3.609
&  $-$0.293
& 3.620 & (120.23) &  186.21
& 3.625 &  3.947
&  $-$0.322
& 3.920 & (181.97) &  269.15
& 3.915 &  4.256
&  $-$0.341 \\
t50g40m2
& 3.320 & ( 63.10) &  100.00
& 3.325 &  3.688
&  $-$0.363
& 3.620 & ( 93.33) &  141.25
& 3.611 &  4.033
&  $-$0.422
& 3.920 & (125.89) &  190.55
& 3.909 &  4.356
&  $-$0.447 \\
t50g30m2
& 3.320 & ( 54.95) &   87.10
& 3.311 &  3.741
&  $-$0.430
& 3.620 & ( 77.62) &  120.23
& 3.610 &  4.161
&  $-$0.551
& 3.920 & (100.00) &  154.88
& 3.911 &  4.533
&  $-$0.622 \\
t50g20m2
& 3.320 & ( 51.29) &   83.18
& 3.326 &  3.805
&  $-$0.479
& 3.620 & ( 70.79) &  112.20
& 3.616 &  4.268
&  $-$0.652
& 3.920 & ( 91.20) &  141.25
& 3.909 &  4.716
&  $-$0.807 \\
t50g10m2
& 3.320 & ( 47.86) &   79.43
& 3.320 &  3.809
&  $-$0.489
& 3.620 & ( 67.61) &  109.65
& 3.630 &  4.321
&  $-$0.691
& 3.920 & ( 87.10) &  138.04
& 3.931 &  4.825
&  $-$0.894 \\
t45g50m2
& 3.320 & (151.36) &  245.47
& 3.316 &  3.686
&  $-$0.370
& 3.620 & (223.87) &  354.81
& 3.615 &  3.998
&  $-$0.383
& 3.920 & (323.59) &  501.19
& 3.914 &  4.300
&  $-$0.386 \\
t45g40m2
& 3.320 & (141.25) &  194.98
& 3.330 &  3.645
&  $-$0.315
& 3.620 & (190.55) &  257.04
& 3.610 &  3.925
&  $-$0.315
& 3.920 & (257.04) &  346.74
& 3.922 &  4.220
&  $-$0.298 \\
t45g30m2
& 3.320 & (109.65) &  151.36
& 3.329 &  3.745
&  $-$0.416
& 3.620 & (138.04) &  190.55
& 3.620 &  4.049
&  $-$0.429
& 3.920 & (173.78) &  234.42
& 3.905 &  4.305
&  $-$0.400 \\
t45g20m2
& 3.320 & ( 95.50) &  134.90
& 3.326 &  3.842
&  $-$0.516
& 3.620 & (117.49) &  165.96
& 3.622 &  4.199
&  $-$0.577
& 3.920 & (141.25) &  199.53
& 3.940 &  4.505
&  $-$0.565 \\
t45g10m2
& 3.320 & ( 87.10) &  125.89
& 3.326 &  3.881
&  $-$0.555
& 3.620 & (107.15) &  154.88
& 3.626 &  4.282
&  $-$0.656
& 3.920 & (128.82) &  181.97
& 3.922 &  4.612
&  $-$0.690 \\
t65g50m3
& 2.320 & (  1.12) &    1.82
& 2.323 &  2.532
&  $-$0.209
& 2.620 & (  2.24) &    3.55
& 2.619 &  2.828
&  $-$0.209
& 2.920 & (  4.37) &    6.92
& 2.916 &  3.126
&  $-$0.210 \\
t65g40m3
& 2.320 & (  1.10) &    1.86
& 2.323 &  2.552
&  $-$0.229
& 2.620 & (  2.19) &    3.63
& 2.619 &  2.849
&  $-$0.230
& 2.920 & (  4.27) &    7.08
& 2.919 &  3.151
&  $-$0.232 \\
t65g30m3
& 2.320 & (  0.98) &    1.74
& 2.318 &  2.570
&  $-$0.252
& 2.620 & (  1.95) &    3.47
& 2.624 &  2.877
&  $-$0.253
& 2.920 & (  3.80) &    6.76
& 2.921 &  3.175
&  $-$0.254 \\
t65g20m3
& 2.320 & (  0.81) &    1.55
& 2.318 &  2.602
&  $-$0.284
& 2.620 & (  1.62) &    3.09
& 2.615 &  2.891
&  $-$0.276
& 2.920 & (  3.31) &    6.31
& 2.924 &  3.201
&  $-$0.277 \\
t65g10m3
& 2.320 & (  0.89) &    1.62
& 2.318 &  2.585
&  $-$0.267
& 2.620 & (  1.74) &    3.24
& 2.625 &  2.893
&  $-$0.268
& 2.920 & (  3.47) &    6.31
& 2.925 &  3.196
&  $-$0.271 \\
t60g50m3
& 2.320 & (  2.04) &    3.31
& 2.318 &  2.537
&  $-$0.219
& 2.620 & (  3.98) &    6.46
& 2.621 &  2.842
&  $-$0.221
& 2.920 & (  7.76) &   12.59
& 2.925 &  3.148
&  $-$0.223 \\
t60g40m3
& 2.320 & (  2.09) &    3.39
& 2.319 &  2.541
&  $-$0.222
& 2.620 & (  4.07) &    6.61
& 2.616 &  2.839
&  $-$0.223
& 2.920 & (  7.94) &   12.88
& 2.925 &  3.153
&  $-$0.228 \\
t60g30m3
& 2.320 & (  1.95) &    3.31
& 2.319 &  2.553
&  $-$0.234
& 2.620 & (  3.89) &    6.46
& 2.617 &  2.850
&  $-$0.233
& 2.920 & (  7.59) &   12.30
& 2.915 &  3.155
&  $-$0.240 \\
t60g20m3
& 2.320 & (  1.82) &    3.16
& 2.316 &  2.565
&  $-$0.249
& 2.620 & (  3.55) &    6.17
& 2.616 &  2.864
&  $-$0.248
& 2.920 & (  6.92) &   12.02
& 2.924 &  3.178
&  $-$0.254 \\
t60g10m3
& 2.320 & (  1.66) &    2.95
& 2.318 &  2.588
&  $-$0.270
& 2.620 & (  3.16) &    5.62
& 2.615 &  2.882
&  $-$0.267
& 2.920 & (  6.17) &   10.96
& 2.919 &  3.190
&  $-$0.271 \\
t55g50m3
& 2.320 & (  3.80) &    6.46
& 2.322 &  2.558
&  $-$0.236
& 2.620 & (  7.41) &   12.30
& 2.617 &  2.853
&  $-$0.236
& 2.920 & ( 14.13) &   23.44
& 2.917 &  3.158
&  $-$0.241 \\
t55g40m3
& 2.320 & (  3.98) &    6.61
& 2.316 &  2.544
&  $-$0.228
& 2.620 & (  7.76) &   12.88
& 2.622 &  2.854
&  $-$0.232
& 2.920 & ( 14.79) &   23.44
& 2.915 &  3.153
&  $-$0.238 \\
t55g30m3
& 2.320 & (  4.07) &    6.61
& 2.320 &  2.542
&  $-$0.222
& 2.620 & (  7.76) &   12.59
& 2.619 &  2.846
&  $-$0.227
& 2.920 & ( 14.79) &   22.91
& 2.917 &  3.152
&  $-$0.235 \\
t55g20m3
& 2.320 & (  3.98) &    6.46
& 2.316 &  2.544
&  $-$0.228
& 2.620 & (  7.59) &   12.30
& 2.618 &  2.850
&  $-$0.232
& 2.920 & ( 14.13) &   22.39
& 2.919 &  3.159
&  $-$0.240 \\
t55g10m3
& 2.320 & (  3.55) &    6.17
& 2.325 &  2.565
&  $-$0.240
& 2.620 & (  6.92) &   11.75
& 2.625 &  2.869
&  $-$0.244
& 2.920 & ( 13.18) &   21.38
& 2.922 &  3.175
&  $-$0.253 \\
t50g50m3
& 2.320 & (  7.08) &   13.18
& 2.325 &  2.602
&  $-$0.277
& 2.620 & ( 13.80) &   25.12
& 2.619 &  2.899
&  $-$0.280
& 2.920 & ( 26.30) &   46.77
& 2.916 &  3.202
&  $-$0.286 \\
t50g40m3
& 2.320 & (  8.32) &   14.45
& 2.325 &  2.575
&  $-$0.250
& 2.620 & ( 15.85) &   26.30
& 2.615 &  2.872
&  $-$0.257
& 2.920 & ( 28.84) &   46.77
& 2.918 &  3.187
&  $-$0.269 \\
t50g30m3
& 2.320 & (  8.71) &   14.13
& 2.320 &  2.556
&  $-$0.236
& 2.620 & ( 16.22) &   25.70
& 2.619 &  2.864
&  $-$0.245
& 2.920 & ( 28.18) &   43.65
& 2.918 &  3.181
&  $-$0.263 \\
t50g20m3
& 2.320 & (  8.51) &   13.49
& 2.322 &  2.546
&  $-$0.224
& 2.620 & ( 15.49) &   24.55
& 2.623 &  2.857
&  $-$0.234
& 2.920 & ( 27.54) &   41.69
& 2.925 &  3.180
&  $-$0.255 \\
t50g10m3
& 2.320 & (  8.13) &   12.88
& 2.319 &  2.546
&  $-$0.227
& 2.620 & ( 14.79) &   23.44
& 2.621 &  2.859
&  $-$0.238
& 2.920 & ( 26.30) &   39.81
& 2.923 &  3.181
&  $-$0.258 \\
t45g50m3
& 2.320 & ( 17.38) &   34.67
& 2.323 &  2.642
&  $-$0.319
& 2.620 & ( 33.11) &   64.57
& 2.620 &  2.950
&  $-$0.330
& 2.920 & ( 60.26) &  120.23
& 2.921 &  3.263
&  $-$0.342 \\
t45g40m3
& 2.320 & ( 18.62) &   35.48
& 2.316 &  2.638
&  $-$0.322
& 2.620 & ( 34.67) &   63.10
& 2.620 &  2.957
&  $-$0.337
& 2.920 & ( 58.88) &  104.71
& 2.927 &  3.291
&  $-$0.364 \\
t45g30m3
& 2.320 & ( 21.88) &   34.67
& 2.324 &  2.566
&  $-$0.242
& 2.620 & ( 38.02) &   57.54
& 2.616 &  2.878
&  $-$0.262
& 2.920 & ( 61.66) &   89.13
& 2.919 &  3.223
&  $-$0.304 \\
t45g20m3
& 2.320 & ( 19.95) &   31.62
& 2.325 &  2.570
&  $-$0.245
& 2.620 & ( 34.67) &   51.29
& 2.616 &  2.884
&  $-$0.268
& 2.920 & ( 53.70) &   77.62
& 2.920 &  3.244
&  $-$0.324 \\
t45g10m3
& 2.320 & ( 17.78) &   28.18
& 2.323 &  2.572
&  $-$0.249
& 2.620 & ( 30.90) &   46.77
& 2.622 &  2.899
&  $-$0.277
& 2.920 & ( 47.86) &   70.79
& 2.924 &  3.260
&  $-$0.336 \\
\hline
\end{tabular}
\end{center}
Note. The case of $\xi$ = 2 $\rm{km} \, {\rm s}^{-1}$ calculation for 
the K I resonance line at 7698.98~$\rm\AA$.
Code ``t$aa$g$bb$m$c$'' denotes the model with 
$T_{\rm eff} = aa \times 100$,
$\log g = bb / 10$, and [Fe/H] (metallicity) = $- c$.
Calculations for each model were made three times corresponding to
three assigned potassium abundances, $A_{i}^{\rm a}$ 
($\equiv$ 5.12 + [Fe/H] + [K/Fe]$_{i}$)
($i$ = 1, 2, 3), where ([K/Fe]$_{1}$, [K/Fe]$_{2}$, [K/Fe]$_{3}$) are
($-0.3$, 0.0, +0.3) for the solar metallicity models ([Fe/H] = 0)
and (+0.2, +0.5, +0.8) for all other metal-deficient models
([Fe/H] = $-1$, $-2$, $-3$).
$W_{i}^{\rm LTE}$ and $W_{i}^{\rm NLTE}$ are the resulting theoretical 
LTE and NLTE equivalent widths (in m$\rm\AA$) corresponding to 
the assigned $A_{i}^{\rm a}$, respectively.
Based on such calculated {\it non-LTE} equivalent width, $W_{i}^{\rm NLTE}$,
two kinds of potassium abundances were inversely computed for the cases of
NLTE ($A_{i}^{\rm N}$) and LTE ($A_{i}^{\rm L}$), from which the non-LTE
abundance correction was eventually evaluated as the difference of
these two, $\Delta_{i}$ ($\equiv A_{i}^{\rm N} - A_{i}^{\rm L}$).
\end{table}

\onecolumn

\begin{figure}
  \begin{center}
    \FigureFile(155mm,155mm){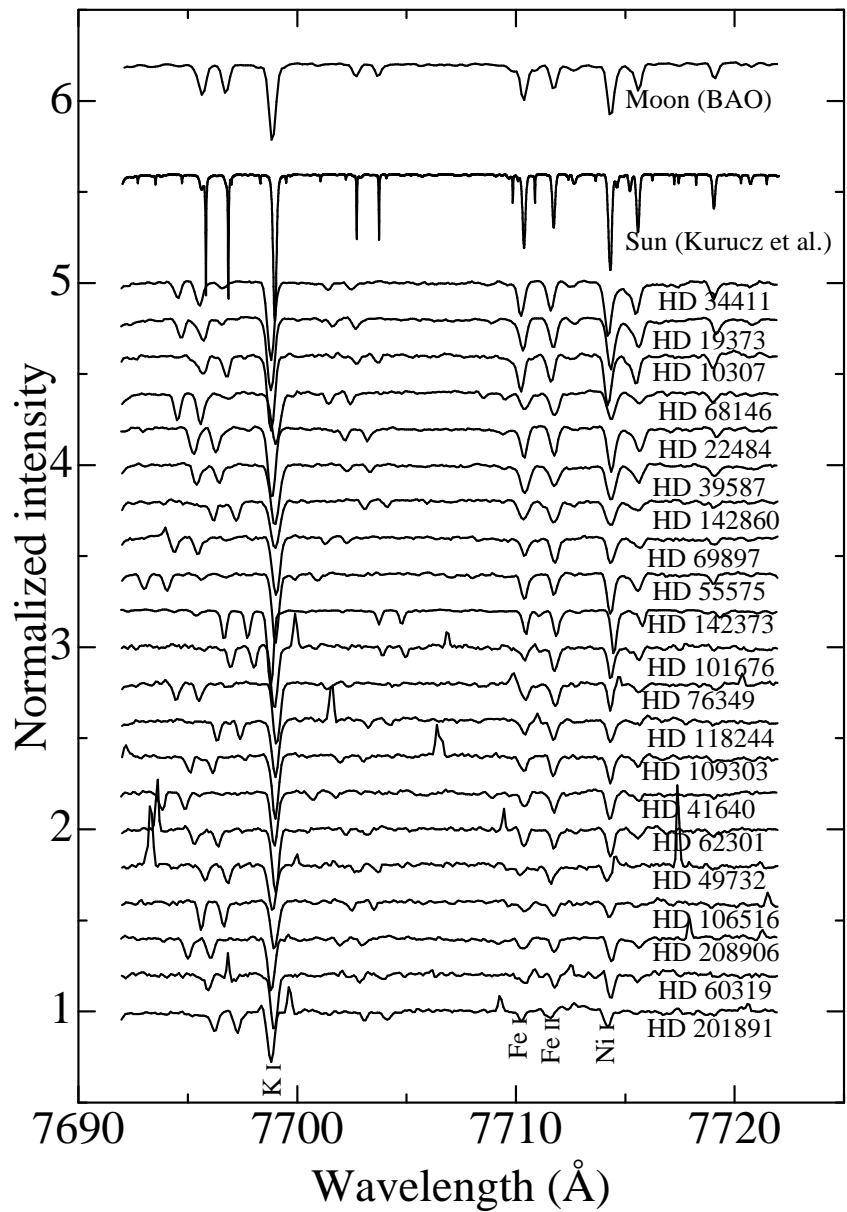}
  \end{center}
\caption{BAO spectra of 21 mildly metal-poor F--G dwarfs
(along with Kurucz et al.'s solar flux spectrum and the BAO Moon 
spectrum; cf. section 2) 
in the 7692--7722~$\rm\AA$ region including the K~{\sc i} 7699 line.
Each spectrum, normalized with respect to the continuum level,
is vertically offset by 0.2 (except for 0.6 for the spectra of 
the Sun and the Moon) relative to the adjacent one.
}
\end{figure}

\begin{figure}
  \begin{center}
    \FigureFile(155mm,155mm){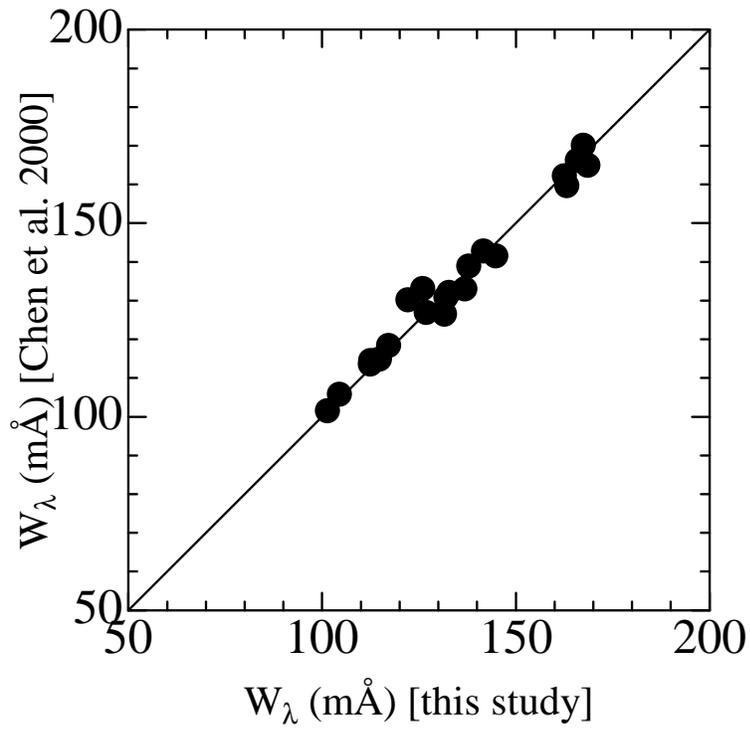}
  \end{center}
\caption{Comparison of the K~{\sc i} 7699 equivalent widths 
of 21 stars used by Chen et al. (2000) with those newly 
remeasured by using the same spectra for this study.
}
\end{figure}

\begin{figure}
  \begin{center}
    \FigureFile(120mm,160mm){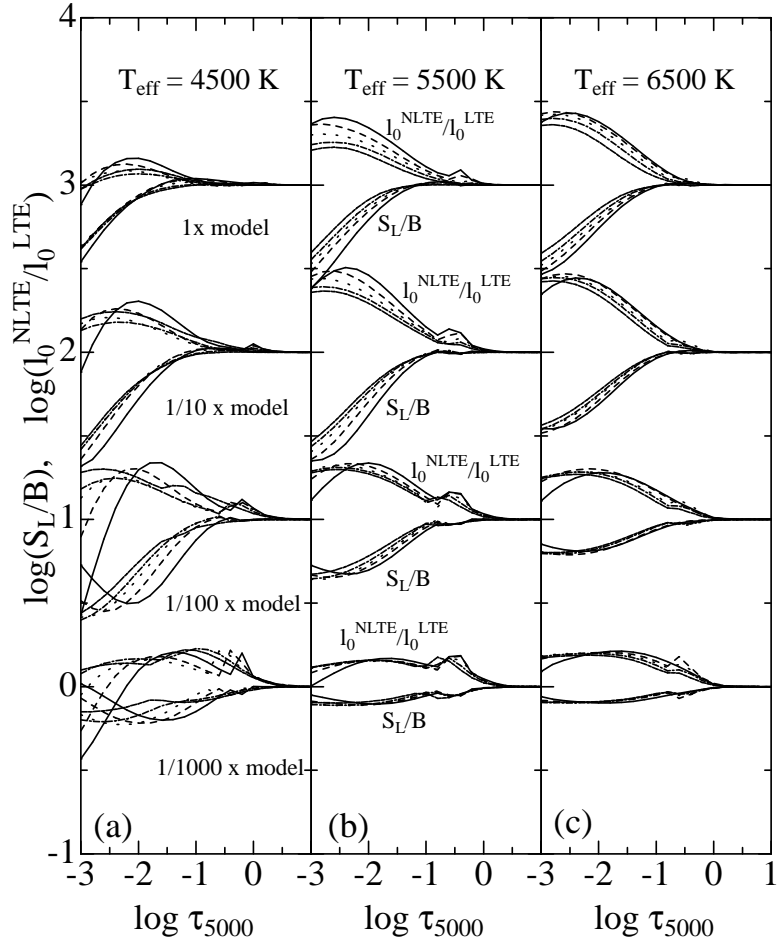}
  \end{center}
\caption{
Ratio of the line source function ($S_{\rm L}$) to the local
Planck function ($B$) and the NLTE-to-LTE line-center opacity
ratio for the 4s~$^{2}$S--4p~$^{2}$P$^{\rm o}$ transition
(corresponding to K~{\sc i} 7698.98 resonance line), which are shown
as functions of the standard continuum optical depth
at 5000 $\rm\AA$ for models of $T_{\rm eff}$ = 4500 K, 5500 K, and 6500 K.
The solid lines, dashed lines, dotted lines, dash-dotted lines, and
dash-double-dotted lines correspond to the results for $\log g$ =
1.0, 2.0, 3.0, 4.0, and 5.0, respectively. Note that the curves are 
vertically offset by an amount of 1.0~dex relative to those of the 
adjacent metallicity ones. 
}
\end{figure}

\begin{figure}
  \begin{center}
    \FigureFile(120mm,120mm){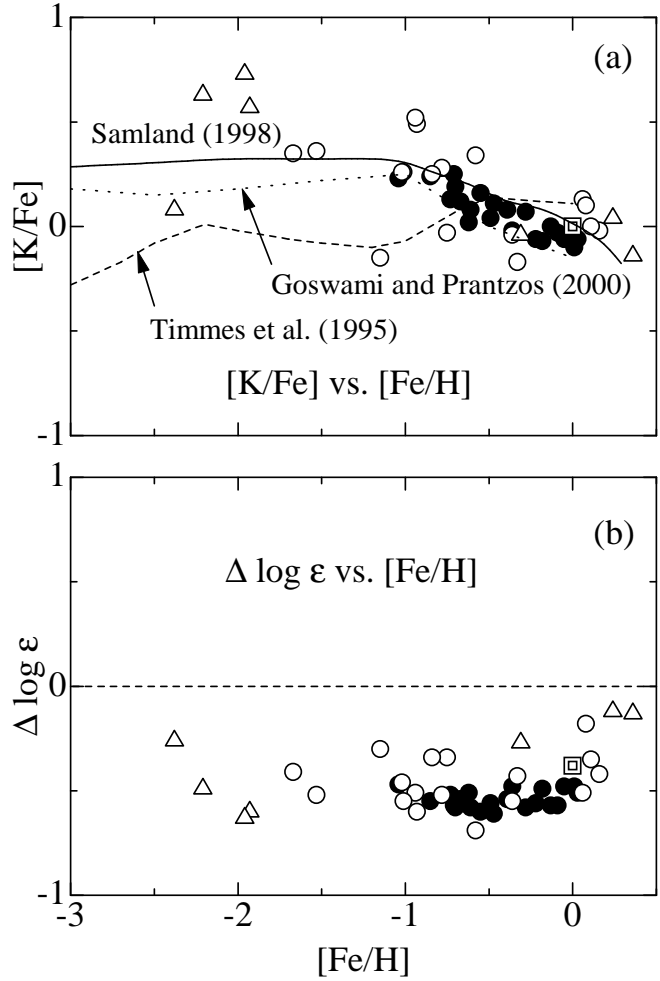}
  \end{center}
\caption{{\scriptsize (a) Resulting [K/Fe] vs. [Fe/H] relation 
constructed from the NLTE abundances of potassium for the program stars. 
Lines show the three kinds of theoretical predictions published so far 
(cf. section 1). Solid line --- Samland (1998) (Salpeter IMF, adjusted 
K yield about twice as large as that of WW95, adjusted Fe yield about 
half as small as that of WW95), dashed line --- Timmes et al. (1995) 
(Salpeter IMF, K yield of WW95, Fe yield of 0.5$\times$WW95), 
dotted line --- Goswami and Prantzos (2000) (their ``Case B'', 
Kroupa et al.'s IMF, Fe yield of 0.5$\times$WW95, K yield of WW95).
The results for the BAO sample (22 mildly metal-poor 
F--G dwarfs) are shown by the filled circles, while those derived
by reanalyzing Gratton and Sneden's (1987b) equivalent widths
for 24 metal-deficient stars are denoted by the open symbols
(open circles --- high-gravity stars with $\log g \ge 3.0$;
 open triangles --- low-gravity stars with $\log g < 3.0$).
 The data for the Sun is indicated by the double square.
(b) The NLTE correction for the K abundance for each star plotted as 
a function of [Fe/H].}
}
\end{figure}

\end{document}